\documentclass[conference]{IEEEtran}
\IEEEoverridecommandlockouts
\usepackage{cite}
\usepackage{amsmath,amssymb,amsfonts}
\usepackage{algorithmic}
\usepackage{graphicx}
\usepackage{textcomp}
\usepackage{xcolor}
\usepackage{textgreek}
\usepackage{fancyhdr}
\def\BibTeX{{\rm B\kern-.05em{\sc i\kern-.025em b}\kern-.08em
    T\kern-.1667em\lower.7ex\hbox{E}\kern-.125emX}}
\begin{document}

\title{Chatbots for Robotic Process Automation: Investigating Perceived Trust and User Satisfaction
%{\footnotesize \textsuperscript{*}Note: Subtitle}
}

\author{\IEEEauthorblockN{Alessandro Casadei}
\IEEEauthorblockA{\textit{Management, Communication \& IT} \\
\textit{MCI | The Entrepreneurial School}\\
Innsbruck, Austria \\
a.casadei@mci4me.at}
\and
\IEEEauthorblockN{Stephan Schl\"{o}gl}
\IEEEauthorblockA{\textit{Management, Communication \& IT} \\
\textit{MCI | The Entrepreneurial School}\\
Innsbruck, Austria \\
stephan.schloegl@mci.edu}
\and
\IEEEauthorblockN{Markus Bergmann}
\IEEEauthorblockA{\textit{Customer Relationship Management} \\
\textit{World-Direct eBusiness Solutions GmbH}\\
Vienna, Austria \\
markus.bergmann@world-direct.at}
}

\maketitle

\thispagestyle{fancy}
%\lfoot{978-1-6654-5238-0/22/\$31.00~\copyright2022 IEEE}
\lfoot{Preprint}
\cfoot{}
\rfoot{}

\begin{abstract}
Driven by ongoing improvements in machine learning, chatbots have increasingly grown from experimental interface prototypes to reliable and robust tools for process automation. Building on these advances, companies have identified various application scenarios, where the automated processing of human language can help foster task efficiency. To this end, the use of chatbots may not only decrease costs, but it is also said to boost user satisfaction. People's intention to use and/or reuse said technology, however, is often dependent on less utilitarian factors. Particularly trust and respective task satisfaction count as relevant usage predictors. In this paper, we thus present work that aims to shed some light on these two variable constructs. We report on an experimental study ($n=277$), investigating four different human-chatbot interaction tasks. 
%All tasks, although attached to the same general theme, exemplify different levels of complexity (i.e., obtain an answer to a simple question, obtain a detailed answer/explanation, solve a simple problem, solve a problem which requires human assistance), which allowed for a broad exploration of user perceptions. 
After each task, participants were asked to complete survey items on perceived trust, perceived task complexity 
%(subdivided into perceived correctness, concreteness, and responsiveness) 
and perceived task satisfaction. 
%(subdivided into perceived effort, appropriateness and fairness). 
Results show that task complexity impacts negatively on both trust and satisfaction. %especially on perceived correctness. 
To this end, higher complexity was associated particularly with those conversations that relied on broad, descriptive chatbot answers, while conversations that span over several short steps were perceived less complex, even when the overall conversation was eventually longer. 
\end{abstract}


\section{Introduction}
%Customer service is a cost centre and, despite increasing volumes and increased customer expectations, the staffing levels are remaining mostly the same. To meet higher expectations and increased demand, 
Today's customer service is increasingly dependent on automation. To this end, chatbots (and other natural language user interfaces) have become a popular way to deal with customers.
%thanks to their natural language user interfaces. 
Respective solutions not only automate current activities but may also help foster customer experiences in that chatbots are available 24/7 and thus significantly reduce waiting times. With information retrieval tasks, they may even outperform search engines~\cite{crutzen2011artificially} and with problem inquiries it has been shown that people are occasionally more open with chatbots than they are with humans, as they do not feel judged when asking trivial questions~\cite{brandtzaeg2018chatbots}.
%For all these motivations, virtual agents are especially well suited for front-office activities.

One way of expanding the capabilities of chatbots beyond their ability to offer a human-like machine interfaces is seen in their potential connection to Robotic Process Automation (RPA). RPA aims to automate highly repetitive business processes. Traditionally, it operates with deterministic inputs and thus, it is only suitable for standardized processes with rule-based decision-making, such as back-office activities. In connection with a chatbot, however, these activities could be triggered more or less directly by a customer. That is, the chatbot's Naural Language Processing (NLP) unit can translate natural language into standardized data, which the RPA engine then uses to (efficiently) complete back-office-related processes.  
%and without human intervention.
%Accordingly, in its classic version, RPA is related to back-office processes.

The work presented in this article aims to provide relevant insights into the potential success factors of this connection between chatbot interaction and RPA. In this, our focus lies on investigating the relationship between people's trust towards chatbots, perceived task complexity as well as people's satisfaction felt when using a chatbot to solve such tasks. 

Our report starts with a discussion of related work in Section~\ref{sec:relatedwork}. Next, we describe the methodological approach of our investigation in Section~\ref{sec:methodology}. Respective results are then presented in Section~\ref{sec:results} and further discussed in Section~\ref{sec:discussion}. Finally, Section~\ref{sec:conclusion} concludes the paper and proposes directions for future work.
%The link between perceived task complexity and the conversation design is also discussed. Finally, the role of user-related factors on reuse intention is analyzed through qualitative methods.
%This article aims to investigate the success factors of a solution that combines chatbots and RPA in end-to-end automation. 
%In fact, chatbots and RPA can be seen as complementary, as AI-enabled chatbots can translate natural language into standardized data thanks to NLP, while RPA can complete back-office-related processes efficiently and without human intervention. 
%The main aspect covered in this work is related to customer perception, since a critical factor for success is the user's reuse of the chatbot, leaving the back-office part of the system for future investigations. In particular, this article focuses on clarifying the relationships between trust, satisfaction and perceived task complexity. The link between perceived task complexity and the conversation design is also discussed. Finally, the role of user-related factors on reuse intention is analyzed through qualitative methods.
%%========================================================================================
\section{Related Work}\label{sec:relatedwork}
%An extract of the literature used in the study will be exposed next, with a special focus on the user perception topic
%\subsection{User perception}
%The perceived social intelligence of chatbots has been investigated ~\cite{mariacher2021investigating} by assessing different human-like attributes, such as authenticity, clarity and empathy ~\cite{article}. 
Previous work has shown that when using chatbots, people seem to care more about task completion efficiency than they care about the interaction experience they are exposed to~\cite{mariacher2021investigating}. However, negative properties of an interaction, such as conversational errors, incomprehension or task complexity, can negatively impact on users' perceived trust in the capabilities of a chatbot and consequently hamper its adoption~\cite{toader2019effect}.

\subsection{Chatbot Trust Perception}
It has been shown that, next to perceived ease of use and perceived usefulness, it is particularly trust which acts as an important predictor for the adoption of chatbot technology~\cite{pillai2020adoption}. As with other products, people's trust perception towards chatbots is often influenced by how reliable and sophisticated an interaction is. Interpretational problems, such as incomprehension leading to long-lasting error recovery routines, significantly lower people's trust perception~\cite{folstad2018makes}. So that in some cases, they refrain from asking complex questions, as they expect the chatbot to not understand and consequently to not be able to correctly answer their demand. Hence, concreteness and answering speed are generally considered to positively influence chatbot trust~\cite{nordheim2018trust}. 

Another issue concerns the use and processing of (often sensitive) data~\cite{hasal2021chatbots}. That is, technology users today increasingly claim their right to privacy preserving data processing which, with respect to chatbots and other AI-driven technologies, can be challenging~\cite{biswas2020privacy}. 

Finally, the risk of potentially undesired outcomes hampers chatbot interactions and may thus be considered a reason for lacking trust~\cite{nordheim2019initial}.

%A research group at the University of Oslo was particularly interested in this topic. 
%First, an exploratory study was carried out on the concept  ~\cite{folstad2018makes}. The findings show the principal issue that affects trust regards interpretational problems. Another factor is that users, in some cases, don't want to ask complex questions because they expect the chatbot to not be able to answer them. The last downside is represented by the desire for human support, which pushes towards a hybrid approach to customer service. 
%Nevertheless, chatbot-related factors were identified as the most influential on user experience, specifically the correctness of the interpretation and, as a secondary factor, the human likeness. However, some factors concerning the service context were proved relevant. 

%The most important one is the brand trust towards the company to which the chatbot belongs, but privacy respect played a role, too.

%Further research confirmed that the chatbot-related factors are the most relevant in terms of trust ~\cite{nordheim2018trust}. 
%Specifically, correctness, concreteness and speed of the answers were the most important features of a trusted chatbot. Consequently, her results served as the basis for assessing trust in the present study.
%A model of trust towards chatbots divided into chatbot-related factors, environment-related factors, user-related factors is also proposed-

\subsection{Chatbot Task Complexity}
The perception as to how complex it is to interact with a chatbot in order to accomplish a given task impacts on the technology's initial use and consequent reuse intention.
%The perceived task complexity is a central aspect of this study, which is focused on assessing its impact on reuse intention. 
Previous work furthermore shows that with respect to chatbots, task complexity takes on a moderating role between the perceived chatbot friendliness and people's trust in the technology~\cite{cheng2021exploring}. In other words, the potentially beneficial effects of chatbot friendliness on customer trust will be less pronounced when a task is perceived to be more complex. Thus, when it comes to seemingly complex activities, customers prefer human over chatbot assistance. On the other hand, it has been shown, that customers are likely to use chatbots for low-complexity activities, since they believe that chatbots may be able to accomplish these tasks better and/or quicker than human operators~\cite{xu2020ai}. 

%On the other hand, users prefer human customer assistance over chatbots for high-complexity activities. 

%Finally, with task difficulty acting as a boundary condition, perceived problem-solving skill mediates the effects of the user's reuse intention.

%This implies that solving customers' problems and raising the chatbot's degree of professionalism should be a top priority for companies, rather than simply providing users with polite, generic responses. 

\subsection{Chatbot User Satisfaction}
Radziwill \& Benton~\cite{radziwill2017evaluating} propose to measure chatbot quality along three characteristics, i.e. effectiveness, efficiency and satisfaction. Similarly, Brandtzaeg \& Folstad~\cite{brandtzaeg2017people} found that the primary reason for people to use chatbots lies in expected productivity gains. Unmet expectations, however, negatively impact on people's service satisfaction, for which chatbot capabilities should be transparently outlined at the beginning of an interaction~\cite{folstad2019chatbots}. Furthermore it has been found that information and service quality are major drivers for user satisfaction~\cite{ji2020effects}, while enjoyment, usefulness and ease of use are predictors for continuance of use~\cite{ashfaq2020chatbot}. To this end, Rossmann et al.~\cite{rossmann2020impact} identified various relevant service dimensions, among which \textit{task effort} and \textit{procedural justice} have been found to be the ones that affect user satisfaction the most.

%%========================================================================================
\section{Methodology}\label{sec:methodology}
Building upon the above presented previous work, the primary goal of our study was to investigate the connection between perceived chatbot task complexity, chatbot trust and chatbot user satisfaction and consequent reuse intention. To do this, we designed an online experiment comprising four different tasks with varying complexity levels (cf. Figs.~\ref{fig:task1}--\ref{fig:task4}). Then we asked study participants to solve these tasks through a chatbot interface we developed using Microsoft's Power Virtual Agent platform\footnote{Online: https://powerplatform.microsoft.com/en-us/power-virtual-agents/ accessed: August 30\textsuperscript{th} 2022}. The chatbot appeared as a web page whose link was included in the experiment introduction.  
Participants were introduced to each task separately, provided by a short textual description. After reading the task description, they were instructed to request assistance from the chatbot, prompting the right task-related conversational flow. The objective was to complete an end-to-end conversation by providing the right data or choosing the correct options to complete the task. 
%These included basic controls, such as answering with a number if the room number was asked. %If a participant could not complete a task, they were required to restart the chat. 
After each task, participants were sent a link to a survey asking them to rate their interaction experience via five point Likert-scale based survey items focusing on chatbot trust and chatbot user satisfaction. 
%through interacting with a chatbot. 
%have been designed and the participants were expected to interact with a chatbot to solve them.

\subsection{Task Design}\label{sec:tasks}
%The primary aim of the experimental part is to acknowledge what contributes to the reuse intention of users towards chatbots, focusing in particular on the task complexity.
%To achieve that, four different tasks, with increasing complexity levels, have been designed and the participants were expected to interact with a chatbot to solve them. 
The goal was to design four realistic tasks whose complexity could be  controlled, allowing for a broad exploration of user perceptions. Consequently, the following task classes were defined: (1) Obtain an answer to a simple question; (2) obtain a detailed answer/explanation; (3) solve a simple problem; and (4) solve a problem which requires human assistance.

%\begin{itemize}
%    \item Obtain an answer to a simple question
%    \item Obtain a detailed answer/explanation
%    \item Solve a simple problem
%    \item Solve a problem which may require human assistance
%\end{itemize}

Next, we decided to give these generic tasks a general theme so that study participants would experience an integrated and authentic problem space. We opted for the tourism domain, in which study participants should put themselves in the role of tourists visiting Vienna, and consequently use a chatbot as a first level customer assistance tool to tackle respective tasks regarding their hotel stay. Thus, the following four tasks were presented to them (note: the task complexity was meant to be ascending for which the task order was the same for all participants):
\vspace{0.5cm}

\subsubsection*{Task 1}
The first task focused on obtaining an answer to a simple question. 
\begin{figure}[h]
    \centering
    \includegraphics[width=0.4\textwidth]{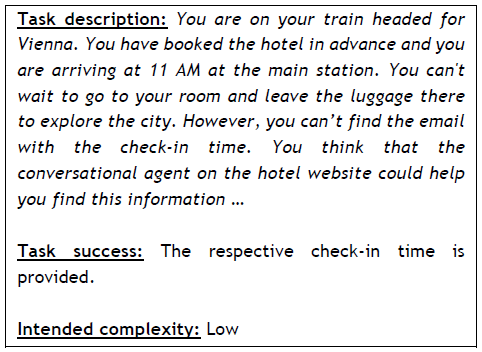}
    \caption{Task 1: Obtaining an answer to a simple question}
    \label{fig:task1}
\end{figure}

\newpage

\subsubsection*{Task 2}
The second task aimed at obtaining a detailed answer and respective description. 
\begin{figure}[h]
    \centering
    \includegraphics[width=0.4\textwidth]{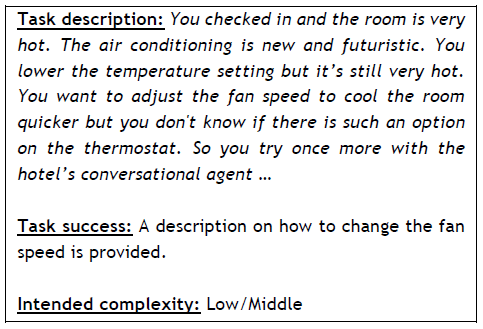}
    \caption{Task 2: Obtaining a detailed answer/explanation}
    \label{fig:task2}
\end{figure}

\subsubsection*{Task 3}
For the third task participants had to solve a concrete but rather simple problem.
\begin{figure}[h]
    \centering
    \includegraphics[width=0.4\textwidth]{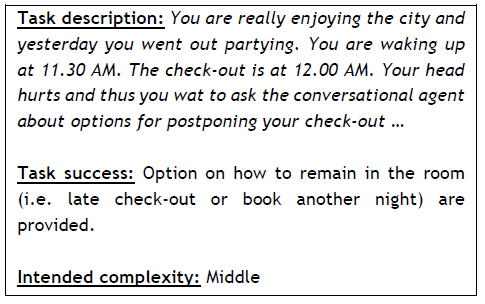}
    \caption{Task 3: Solving a simple problem}
    \label{fig:task3}
\end{figure}

\subsubsection*{Task 4}
Finally, to solve the fourth task participants had to ask for human assistance.
\begin{figure}[h]
    \centering
    \includegraphics[width=0.4\textwidth]{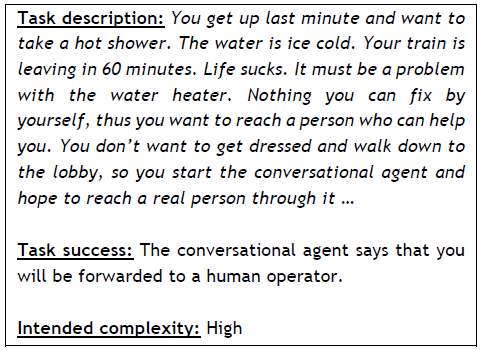}
    \caption{Task 4: Solving a problem which requires human assistance}
    \label{fig:task4}
\end{figure}

%They were then confronted with four different problems/tasks regarding their hotel, for which they were asked to use the hotel chatbot as a customer assistance tool.   
%during this journey they will have four problems to solve regarding their stay in a hotel. 
%The hotel website provides them with a chatbot to help them. 

%therefore a context is provided to the participants.

%All tasks, although attached to the same general theme, exemplify different levels of complexity (i.e., obtain an answer to a simple question, obtain a detailed answer/explanation, solve a simple problem, solve a problem which requires human assistance), 

%The experiment is designed to represent a realistic situation; therefore a context is provided to the participants. 
%The participants are asked to identify themselves as tourists visiting Vienna for the weekend and during this journey they will have four problems to solve regarding their stay in a hotel. The hotel website provides them with a chatbot to help them. After each task completion, a questionnaire was required to be filled in to assess the participant’s experience. 

\subsection{Sampling Frame and Procedure}
We aimed for a generic, gender-balanced study sample. No age or technology experience requirements were set so as to allow for a broad spectrum of potential participants. Language-wise, we targeted mainly native (or close to native) speakers of English. 

Most study participants were recruited via Prolific\footnote{Online: https://www.prolific.co/ accessed: August 25\textsuperscript{th} 2022}, a micro-worker platform that previous work has shown to be trustworthy in terms of participants’ attention, comprehension, honesty and reliability~\cite{peer2021data}. Additionally we used social networks (i.e., Linkedin and WhatsApp) as well as direct email to attract potential study participants. 
A total of $n=277$ participants (76\% English native speakers of whom 69\% lived in the UK at the time of the study) completed the entire study, i.e. all of the four tasks and the respective survey items.
The average age of participants was $38.5$ ($min=19; max=76; SD=13.8$). 
%The only recruitment criterion was sex, in order to achieve an equal representation of male and female users in the sample, 
%and country, to include as many English native speakers as possible. 

\subsection{Post-Task Survey}
\begin{figure*}
    \centering
    \includegraphics[width=0.75\textwidth]{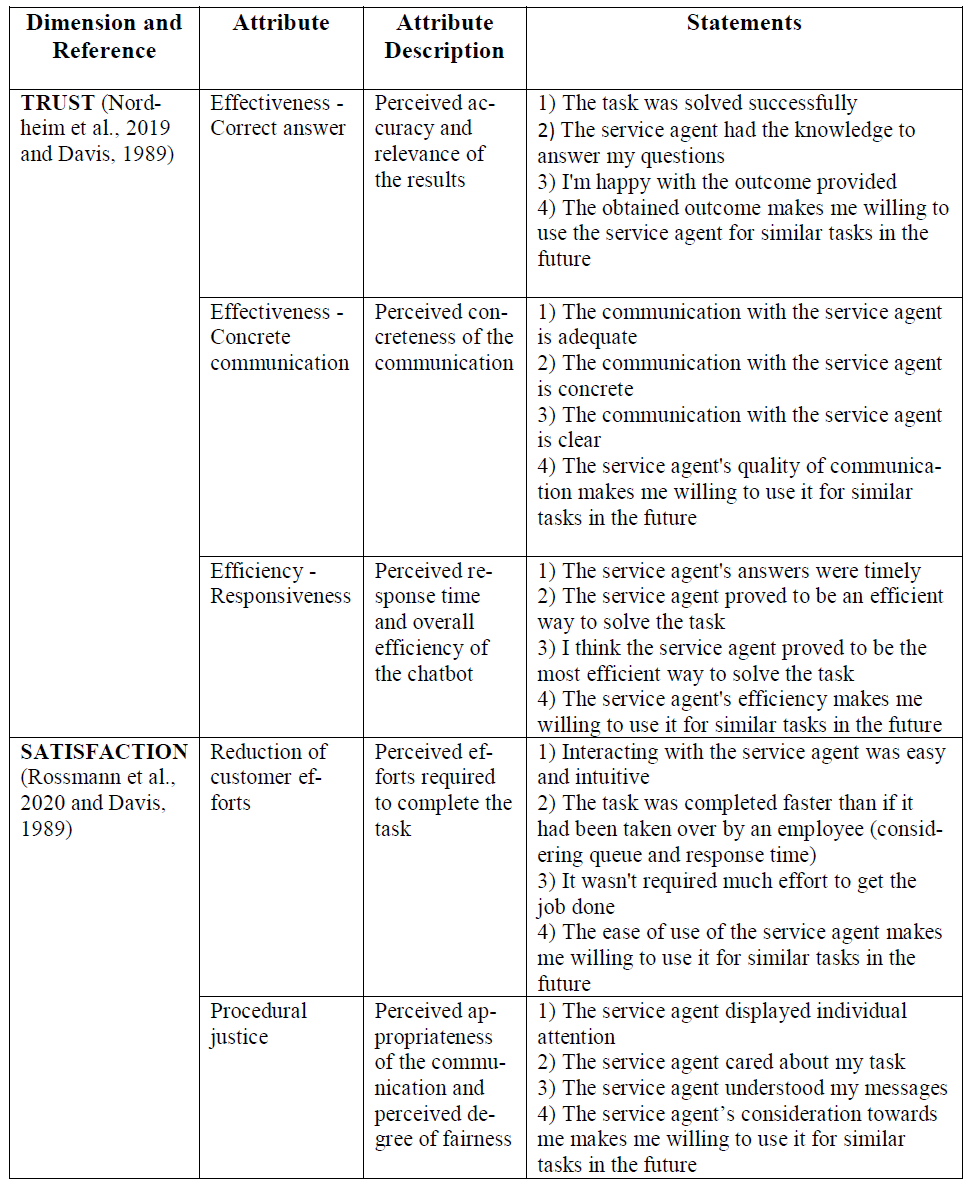}
    \caption{Survey items focusing on chatbot trust and user satisfaction}
    \label{fig:questionnaire}
\end{figure*}

As outlined earlier, participants had to use our Power Virtual Agent chatbot to complete the above outlined tasks, after each of which they were asked to complete survey items assessing their interaction experience. Items focused on perceived task complexity (note: we consciously designed tasks that would represent four different levels of complexity, cf. Section~\ref{sec:tasks}, Figs.~\ref{fig:task1}-\ref{fig:task4}), trust and user satisfaction.
%The two main dimensions used to evaluate the user experience are trust and satisfaction. 
The latter two dimensions have shown to be considerably relevant in predicting reuse intention~\cite{nordheim2018trust,rossmann2020impact,ji2020effects}. To this end, trust was found to be mainly rooted in utilitarian factors, and thus to play a rather complementary role to satisfaction. Our survey aimed to measure it through participants' perceptions concerning the chatbot's \textit{answer correctness}, \textit{communication effectiveness} and \textit{responsiveness}~\cite{nordheim2019initial}. As for satisfaction, however, we focused on the already discussed insights of Rossmann et al.~\cite{rossmann2020impact} and thus used measures of perceived \textit{task effort} and \textit{procedural justice}~\cite{rossmann2020impact}. A copy of the respective survey items for both trust and satisfaction is depicted in Fig.~\ref{fig:questionnaire}.

Finally, after completing all four tasks, participants were asked to order them concerning their perceived task complexity.
%to research the impact of the perceived task complexity on reuse intention.
Furthermore, they had to answer question items asking about their general experience with technology, their previous experience with chatbots and their experience with (other) online services, before being confronted with an optional open question asking about general feedback concerning the accomplished tasks.
%The factors used to represent the user’s trust level towards the chatbot were correct answer, effectiveness and responsiveness .

%The reasons for this design lie in the fact that satisfaction alone does not always lead to higher reuse. The literature suggests that highly friendly and anthropomorphic chatbots are expected to score high in satisfaction, however, since the users are distracted from the tasks they are trying to solve, the perceived usefulness may decrease, leading to lower reuse intention \cite{gumucs2021effect}

%Consequently, trust, which is mainly rooted in utilitarian factors, plays a complementary role to satisfaction. 
%The factors used to represent the user’s trust level towards the chatbot were correct answer, effectiveness and responsiveness \cite{nordheim2019initial}. 

%On the other hand, satisfaction  is represented by reduction of customer efforts and procedural justice \cite{rossmann2020impact}. The factors above have proven to be the most critical predictors of trust and satisfaction, respectively.
%%========================================================================================
\section{Results}\label{sec:results}
%As a result, approximately 76\% of the participants were from USA or UK.
%231 participants were recruited on Prolific, which is one of the most trustworthy platforms in terms of participants’ attention, comprehension, honesty and reliability \cite{peer2021data}. 
%Another pool of 51 participants was recruited among acquaintances contacted on social networks (Linkedin and Whatsapp) or by email. 
%After the exclusion from the Prolific sample of responses from the same participant and responses with variance equal to zero in each of the four tasks, a total number of 260 answers were recorded for the first task, 240 for the second, 247 for the third and 230 for the fourth.
%\subsection{Construct Reliability}
Looking at the two key measures chatbot trust and user satisfaction, the collected data shows high reliability for all constructs (Cronbach's \textalpha\ $>0.90$).

\subsection{Perceived Task Complexity}
Results show that on a ranking scale from $1=easiest\:task$ to $4=hardest\:task$, participants perceived the first task to have been the easiest (Rank: $Mean=1.42; Median=1.00, Mode=1.00, SD=0.83$), and the fourth task the hardest one to accomplish (Rank: $Mean=3.82; Median=4.00; Mode=4.00, SD=0.55$). This aligns with our intention to present to them tasks in ascending order of complexity. As for tasks two and three, however, participants had a different impression. That is, the second task was felt to be more complex (Rank: $Mean=2.47; Median=3.00; Mode=3.00; SD=0.76$) than the third task (Rank: $Mean=2.13; Median=2.00; Mode=2.00; SD=0.74$). A reason for this reversed complexity perception may be found in the rather long chatbot answer, which was provided as a solution to the second task. Here, participants were asked to use the chatbot to investigate on how to lower the room's temperature. Once the chatbot had recognized this intention, the following instruction was presented:

%the second task as the third most complex (mean=2,47, median=3, mode=3, st.dev.=0,76), the third task as the second most complex (mean=2,13, median=2, mode=2, st.dev.=0,74) and the last task as the most complex . %The tasks were designed to be in ascending order of complexity. Accordingly, the first and the fourth tasks are in line with the expected results, however, the second and the third task have an inverted complexity compared to the expected ones.

%By looking in-depth at the design of these two tasks, it is possible to acknowledge the cause of this unexpected outcome.
\vspace{0.2cm}
\textit{``To make quick temperature adjustments, press the Temperature Up or Temperature Down arrows on the thermostat display. If the mode Heat or Cool is on, you can press the Fan indicator to switch between the different fan modes. Tip: If you have set a schedule for your thermostat, manual changes temporarily override the schedule. The temperature you requested holds until the next scheduled change time, such as from Home to Sleep.''}

The third task, on the other hand, consisted of several steps in order to confirm a late check-out or to book the room for an extra day. Here a participant was required to follow the following three steps so as to complete the task:

\begin{enumerate}
    \item \textit{State your room number;}
    \item \textit{Inquire whether a late check-out or a booking extension is available;} 
    \item \textit{Confirm the late check-out or complete the reservation for an extra day;}
\end{enumerate}

%1.	Write the room number. \\
%2.	Acknowledge that both the late check-out and the booking were available and choose between them. \\
%3.	Confirm the late check-out or complete the reservation for extra days. \\

The higher number of steps and respective decisions that had to be made were supposed to lead to a higher level of perceived complexity. Participants, however, seem to have felt that reading and interpreting a long message, as it was provided in Task 2, is cognitively more demanding and consequently more complex than following a (small) number of relatively simple steps, as was required for Task 3.

%However, the results refute this assumption. Participants seem to have required less cognitive effort to follow many small steps compared to reading a relatively long message to solve a task. 

%This is reasonable as long as each step includes short and clear messages and simple decisions. 
%As a result, the tasks that can be completed with several small steps are likely to be more suited for chatbot automation compared to the tasks which imply a descriptive solution. This aspect can be considered complementary to the current literature. Research on FITD (Foot-in-the-door) technique suggests that if the chatbot provides an easy step at the beginning of the task the user will be more prone to trust it for the next steps \cite{adam2021ai}. 

%Overall, the best case would be a task consisting exclusively of easy steps, however, the purpose of chatbots is to solve different kinds of requests and the more complex ones may not be suited for this type of design. In this case, including the easiest steps at the beginning and adding the ambiguous steps later on in the conversation would be the choice leading to higher user compliance, as the FITD technique indicates. \\

\subsection{The Impact of Task Complexity}
A Pearson correlation was used to asses the impact of task design on human-chatbot interaction.
The survey data confirms previous work~\cite{cheng2021exploring} in that it points to a significant negative relationship between perceived task complexity and all the measured trust constructs ($p<0.001$): i.e. \textit{answer correctness}: $-0.622$; \textit{communication effectiveness}: $-0.528$;  \textit{responsiveness}: $-0.406$. 

Similarly, the data suggests a significant negative connection between \textit{perceived task complexity} and participants' satisfaction with the chatbot ($p<0.001$): \textit{task effort}: $-0.500$; \textit{procedural justice}: $-0.445$. Here, the correlation with \textit{procedural justice} particularly supports Cheng et al.'s claim that the perceived friendliness of a chatbot may significantly decrease with increasing task complexity~\cite{cheng2021exploring}. 

\subsection{Qualitative Findings}
%After they had completed all tasks, participants were given the opportunity to comment on whether in a similar future task setting they would use a chabot or rather prefer the interaction with a human operator.
After participants had completed all tasks, they were given the opportunity to comment on whether they would use the chatbot again in a similar future task setting. An inductive approach was then used to classify the comments in three categories: preference for chatbots, preference for human assistance and preference depending on the task complexity.
%preference for reusing such an AI-based system in a similar setting over a human operator. 
A total of $37$ responses were collected. 
%and clustered int three specific categories. 
%\subsubsection{Preference for Human Assistance}
Of these $37$ responses, $10$ respondents clearly underlined their preference for being helped by a human operator. They did not raise any concerns regarding the chatbot's ability to help or communicate. Rather, they expressed their desire for a more human touch, generally disliking the idea of chatting to or with an artificial entity. 
%answers based on subjective reasons, 10 preferred to talk with real employees. 
%This category of respondents emphasised the importance of human touch. Their statements are not concerned with the chatbot's lack of ability to communicate like a human, but they rather disliked interacting with a chatbot \emph{a priori}, at least compared to human conversations. 
To this end, the simple awareness that one is interacting with a chatbot seems to be  enough to make users feel uncomfortable. As one of the respondents exemplifies:
\textit{``[…] I understand that chatbots save time (and money!) to their company but I still prefer a proper person to have a conversation with.''} 

%If a part of users tends to reject chatbots regardless of their design and effectiveness, a change of focus from chatbot design to the customer may be necessary. In other words, it would be beneficial to ensure that the customer base is willing to interact with chatbots at all before planning customer service automation. Specifically, certain criteria need to be established to identify the right psychological attributes so that users do not feel uncomfortable when conversing with a chatbot. 
%A possible solution to adapt the chatbot to the users who require a human touch seems to be to increase the human traits of the chatbot. However, when this happens, the users’ sense of uneasiness is likely to increase, due to the uncanny valley effect. This effect describes the connection between the human-like appearance of a robotic item and the emotional reaction it inspires. When the human traits are too high, the emotional reaction is negative and repulsive. As a consequence, the best choice would be to simply provide human support.

%\subsubsection{Preference for Chatbot Interaction}
Contrary to this antipathy for language-based human-technology interaction, $9$ respondents expressed their clear preference for using a chatbot over talking to a human operator. 
%This category justified their opinion on the basis of a subjective and psychological attitude similar to the category above. 
Here, the most common reason was given in the desire to avoid human conversation altogether. As stated by one respondent: \textit{``chatbots are a great solution especially for the younger generation with anxieties for making a phone call''}. Previous work supports this statement, in that it has shown that people with a certain level of social phobia are more prone to engaging with (anthropomorphic) chatbots than with human beings~\cite{jin2021consumers}.
%and the consequent use of chatbots as a tool to achieve this avoidance. 
%This result strengthens the findings on this topic present in the literature, which acknowledges that users with high social phobia are more prone to like anthropomorphism in chatbots and have a higher intention of reuse compared to users with low social phobia \cite{jin2021consumers}. 
%Although this study is not concerned with social phobia, it is clear that 9 participants felt uncomfortable with the idea of talking with a person to solve their problems and, thus, appreciate chatbots more. 
%A participant’s answer exemplifies this concept effectively: \\
%\emph{“chatbots are great solution especially for younger generation with anxiety of making a phone call”}
%\subsubsection{Preference depending on the task complexity}

The remaining $18$ responses did neither show a clear preference for nor against chabtbot interaction. It was rather pointed out that the choice would depend on the type of task to be accomplished. As one respondent puts it: \textit{``I think I would be ready to reuse the chatbot for simpler and more straightforward tasks. However, I feel in more complex or less common scenarios, like the fourth task, I would probably prefer speaking to a member of staff instead.''}
%participants stated that their preference for human employees or chatbots would depend on the task to be accomplished. 
%This category of respondents is likely to have neither a relatively high level of social discomfort nor a strong need for human touch. 
%Therefore, their decision is made on a more objective basis compared to the other categories, as their 
In other words, the decision to chose one over the other seems to be grounded in the expected effectiveness and efficiency with which the operator of choice (i.e., chatbot or human) may solve a given problem. 
%This category of customers is directly related to the quantitative findings of this study since it is the only one that values the chatbot-related attributes.
Looking at the statistical data, it can further be seen that respective respondents were mostly satisfied with the way the chatbot helped them solve the first three tasks, while for the fourth task, satisfaction was consistently lower. This also confirms previous work in that users' preferences for chatbot use is higher for low-complexity tasks~\cite{xu2020ai}.% than for high-complexity tasks~\cite{xu2020ai}. 

%But they also pointed out that a chatbot is not an effective tool to solve the fourth one. 
%The aim of the fourth task was, indeed, to get in touch with an employee through the chatbot, since the chatbot could not solve the problem itself. These results reinforce the quantitative findings, as the first three tasks were associated with very similar scores in all dimensions, while the fourth task was linked to significantly lower scores. 

%The example below effectively expresses the role of perceived task complexity for this category of users:\\

Lastly, our data shows that existing technology experience is positively connected to chabot trust and user satisfaction, supporting the assumption that also natural language based human-computer interaction may need to be learned. 

%%========================================================================================

\section{Discussion}\label{sec:discussion}
%Text...
%\subsection{Conversation attraction: human or chatbot?}
Literature states that missing user focus is one of the main reasons for chatbot applications to fail~\cite{brandtzaeg2018chatbots}. A comprehensive understanding of potential chatbot users, their perceptions, attitudes as well as preferences thus counts as a necessary prerequisite for the respective technology to be adopted~\cite{pillai2020adoption}. Still, today we see a prevailing preference for human-human over human-chatbot interaction~\cite{Lei2021-rc}. While our results confirm this merit for human operators, they also point to a specific group of users who see chatbot interaction as a welcoming possibility to avoid human contact. In this, our findings support the work of M\"{u}ller et al.~\cite{muller2019chatbot}, who found that specific personality traits and psychological characteristics act as significant predictors for chabtot acceptance.  
%Personality may thus also be a valid predictor for whether people prefer a human or a chatbot to help them accomplish a task. In other words, it may be necessary to evaluate a target groups personal traits and psychological attributes before attempting any kind of task automation.

With respect to our research agenda, results show that perceived task complexity influences people's preferences for or against chatbot use. Contrary to our expectations, however, complexity was not dependent on the number of steps needed to achieve a task. Rather, it was shown that a multi-step task, requiring several small steps, is often perceived to be less complicated than a single-step task, which may in turn require greater text understanding. Such also aligns with previous work, suggesting that if a chatbot offers a simple step at the beginning of an interaction, users are more likely to trust the chatbot's ability to also solve the subsequent steps~\cite{adam2021ai}. 

Finally, our findings confirm the work of Cheng et al.~\cite{cheng2021exploring} in that they generally point to a significant negative correlation between perceived task complexity and people's trust in a chatbot's ability. To this end, it is assumed that more complicated tasks require more user effort, for which also the perceived usefulness of such a system decreases with increasing task complexity. In accordance with Davis et al.~\cite{davis1989user} this may further hamper the overall acceptance and consequent adoption of the chabot.

\section{Conclusion and Future Work}\label{sec:conclusion}
Our study explored the use of a chatbot interface in combination with RPA. Data shows that perceived task complexity impacts on people's trust in and preference for using such an interface in a given problem setting. Consequently,  tasks which require a seemingly higher cognitive understanding (particularly on the technology side) are less likely to be chosen for a chatbot-driven solution approach.  
%The ideal tasks appear to be the ones related to a clear and concise answer or the ones which perform an action. On the other hand, the tasks leading to a descriptive answer or that cannot be immediately and easily solved by the chatbot have proved to lead to low reuse intention. 
%Similarly, the conversation designs associated with higher reuse intention are the ones which require lower cognitive effort, i.e. those composed of simple messages and that guide the user by providing him straightforward choices. If this kind of design is followed, the user will endure a relatively long conversation. 
%The third topic regards the recognition of broader causes leading to a failure of a chatbot solution. 
Results, however, also show that users' personality and psychological traits influence the preference for or against chatbot engagement. 

As for future research directions, we propose to focus on three aspects. First, since our findings suggest a clear relationship between task complexity and potential chatbot use, it seems relevant to better understand the characteristics which make a task seem difficult. Second, building upon the assumption that for seemingly difficult tasks the use of a chatbot may not even be considered, it seems relevant to investigate ways to lower this entry barrier. And finally, we suggest for future work to also focus on improving the ways chatbot engagement may be adapted to individual user preferences, particularly with respect to personality traits and characteristics.

%which underlines the need for additional studies in this field. 

%Consequently, adopting a customer-centric approach seems essential to identify both the right tasks as well as the right users for chatbot-driven RPA.
%and replicating the methodology used in this study could help to test chatbot solutions from a customer perspective.

%%========================================================================================

\bibliography{ichms2022}
\bibliographystyle{IEEEtran}

%\begin{thebibliography}{00}
%\bibitem{b1} G. Eason, B. Noble, and I. N. Sneddon, ``On certain integrals of Lipschitz-Hankel type involving products of Bessel functions,'' Phil. Trans. Roy. Soc. London, vol. A247, pp. 529--551, April 1955.
%\bibitem{b2} J. Clerk Maxwell, A Treatise on Electricity and Magnetism, 3rd ed., vol. 2. Oxford: Clarendon, 1892, pp.68--73.
%\bibitem{b3} I. S. Jacobs and C. P. Bean, ``Fine particles, thin films and exchange anisotropy,'' in Magnetism, vol. III, G. T. Rado and H. Suhl, Eds. New York: Academic, 1963, pp. 271--350.
%\bibitem{b4} K. Elissa, ``Title of paper if known,'' unpublished.
%\bibitem{b5} R. Nicole, ``Title of paper with only first word capitalized,'' J. Name Stand. Abbrev., in press.
%\bibitem{b6} Y. Yorozu, M. Hirano, K. Oka, and Y. Tagawa, ``Electron spectroscopy studies on magneto-optical media and plastic substrate interface,'' IEEE Transl. J. Magn. Japan, vol. 2, pp. 740--741, August 1987 [Digests 9th Annual Conf. Magnetics Japan, p. 301, 1982].
%\bibitem{b7} M. Young, The Technical Writer's Handbook. Mill Valley, CA: University Science, 1989.
%\end{thebibliography}
%\vspace{12pt}
%\color{red}
%IEEE conference templates contain guidance text for composing and formatting conference papers. Please ensure that all template text is removed from your conference paper prior to submission to the conference. Failure to remove the template text from your paper may result in your paper not being published.

\end{document}